\documentclass[pra,twocolumn,showpacs,superscriptaddress]{revtex4}

\usepackage{graphicx}
\usepackage{latexsym}

\newcommand{\tr}{\mbox{tr}}
\newcommand{\la}{\langle}
\newcommand{\ra}{\rangle}

\begin{document}
\bibliographystyle{apsrev}

\title{A simple formula for the average gate fidelity of a quantum
  dynamical operation}

\date{\today}

\author{Michael A. Nielsen}
\affiliation{Centre for Quantum Computer Technology and
Department of Physics, University of Queensland, Brisbane,
Queensland 4072, Australia} 

\begin{abstract}
  This note presents a simple formula for the average fidelity
  between a unitary quantum gate and a general quantum operation on a
  {\em qudit}, generalizing the formula for qubits found by Bowdrey
  {\em et al} [Phys.  Lett. A {\bf 294}, 258 (2002)].  This formula
  may be useful for experimental determination of average gate
  fidelity.  We also give a simplified proof of a formula due to
  Horodecki {\em et al} [Phys. Rev. A {\bf 60}, 1888 (1999)],
  connecting average gate fidelity to entanglement fidelity.
\end{abstract}

\pacs{03.67.-a,03.65.-w,89.70.+c}

\maketitle

%
%
Characterizing the quality of quantum channels and quantum gates is a
central task of quantum computation and quantum
information~\cite{Nielsen00a}.  The purpose of this note is to present
a simple formula for the {\em average fidelity} of a quantum channel
or quantum gate.

%
%
The average fidelity of a quantum channel described by a
trace-preserving quantum operation ${\cal E}$~\cite{Nielsen00a} is
defined by
\begin{eqnarray}
  \overline{F}({\cal E}) \equiv \int d\psi \la \psi|{\cal E}(\psi) |\psi\ra,
\end{eqnarray}
where the integral is over the uniform (Haar) measure $d\psi$ on state
space, normalized so $\int d\psi = 1$.  We assume ${\cal E}$ acts on a
qudit, that is, a $d$-dimensional quantum system, with $d$ finite.  We
use the notational convention that $\psi$ indicates either $|\psi\ra$
or $|\psi\ra \la \psi|$, with the meaning determined by context.
$\overline{F}({\cal E})$ quantifies how well ${\cal E}$ preserves
quantum information, with values close to one indicating information
is preserved well, while values close to zero indicate poor
preservation.  $\overline{F}({\cal E})$ may be extended to a measure of
how well ${\cal E}$ approximates a quantum gate, $U$,
\begin{eqnarray}
  \overline{F}({\cal E},U) \equiv \int d\psi \la \psi|U^{\dagger} 
{\cal E}(\psi) U |\psi\ra.
\end{eqnarray}
Note that $F({\cal E},U) = 1$ if and only if ${\cal E}$ implements $U$
perfectly, while lower values indicate that ${\cal E}$ is a noisy
implementation of $U$.  Note that $\overline{F}({\cal E},U) =
\overline{F}({\cal U^{\dagger}} \circ {\cal E})$, where ${\cal
  U^{\dagger}}(\rho) \equiv U^{\dagger} \rho U$, and $\circ$ denotes
composition.

%
%
The paper is structured as follows.  First, we state and provide a
simple proof of a result of M.,~P. and R.~Horodecki connecting
$\overline{F}({\cal E})$ to the entanglement fidelity introduced
in~\cite{Schumacher96a}.  We then use the Horodecki's result to obtain
an explicit formula for the average fidelity $\overline{F}({\cal
  E},U)$.  The paper concludes with a discussion of how the formula
for $\overline{F}({\cal E},U)$ may be useful for experimentally
quantifying the quality of quantum gates and quantum channels.

%
%
The present work is a development of the paper of Bowdrey {\em et
  al}~\cite{Bowdrey02a}, who obtained a simple formula for
$\overline{F}({\cal E},U)$ when ${\cal E}$ and $U$ act on qubits. This
paper generalizes to the case where ${\cal E}$ and $U$ act on qudits.
Related results were also obtained by Fortunato \emph{et
  al}~\cite{Fortunato02a,Fortunato02b} who found a simple and
experimentally useful formula for the entanglement fidelity;
\cite{Fortunato02b} had also rediscovered the connection between
average fidelity and entanglement fidelity proved
in~\cite{Horodecki99c}, for the special case of a qubit, thus enabling
them to recover the results of~\cite{Bowdrey02a}.

%
%
To define entanglement fidelity, imagine ${\cal E}$ acts on one half
of a maximally entangled state.  That is, if ${\cal E}$ acts on a
qudit labelled $Q$, then imagine another qudit, $R$, with $RQ$
initially in the maximally entangled state $\phi$.  The entanglement
fidelity is defined to be the overlap between $\phi$ before and after
the application of ${\cal E}$~\footnote{Our definition is a special
  case of~\cite{Schumacher96a}, which also considered non-maximally
  entangled states of $RQ$.}, $F_{\rm e}({\cal E}) \equiv \la \phi|
({\cal I} \otimes {\cal E})(\phi) |\phi\rangle,$ where ${\cal I}$
denotes the identity operation on system $R$.  The entanglement
fidelity is thus a measure of how well entanglement with other systems
is preserved by the action of ${\cal E}$.  Using the fact that any two
maximally entangled states on $RQ$ are related by a unitary on system
$R$ alone, it follows that the value of the entanglement fidelity does
not depend upon which maximally entangled state $\phi$ between $R$ and
$Q$ is used in the definition of entanglement
fidelity~\cite{Schumacher96a}.

%
%
M.,~P., and R.~Horodecki have presented a beautiful
formula~\cite{Horodecki99c} connecting $\overline{F}({\cal E})$ to
$F_{\rm e}({\cal E})$:
\begin{eqnarray}
  \label{eq:Horodecki}
  \overline{F}({\cal E}) = \frac{d F_{\rm e}({\cal E})+1}{d+1}.
\end{eqnarray}
We now give a proof of Eq.~(\ref{eq:Horodecki}), substantially
simplifying the proof in~\cite{Horodecki99c}.  The first step is to
define a new, ``twirled'' operation ${\cal E}_T$, ${\cal E}_T(\rho)
\equiv \int dU U^{\dagger} {\cal E}(U\rho U^{\dagger}) U$, where the
integral is over the normalized uniform (Haar) measure $dU$ on the
space of $d \times d$ unitary matrices.  Note that ${\cal E}_T$ is a
trace-preserving quantum operation.  Next, we argue that twirling does
not change the average fidelity, since
\begin{eqnarray}
\overline{F}({\cal E}_T) & = & \int d\psi \int dU \la \psi |U^{\dagger}
{\cal E}(U\psi U^{\dagger}) U|\psi\ra \\
 & = & \int dU \int d\psi \la \psi |U^{\dagger}
{\cal E}(U\psi U^{\dagger}) U|\psi\ra \label{eq:inter1} \\
 & = & \int dU \overline{F}({\cal E}) = \overline{F}({\cal E}), 
\label{eq:inter2} 
\end{eqnarray}
where Eq.~(\ref{eq:inter2}) follows from Eq.~(\ref{eq:inter1}) by the
change of variables $|\psi'\ra \equiv U|\psi\ra$.  A similar argument
shows that twirling does not change the entanglement fidelity, for if
$\phi$ was the maximally entangled state of $RQ$ then~\footnote{Note
  that $U$ and ${\cal E}$ act on system $Q$ alone in these
  expressions, with the identity action on $R$ implicit.}
\begin{eqnarray}
F_{\rm e}({\cal E}_T) & = & \int dU \la \phi |U^{\dagger}
 {\cal E} \left( U\phi U^{\dagger}\right) U|\phi\ra \\
 & = & \int dU F_{\rm e}({\cal E}) = F_{\rm e}({\cal E}),
\end{eqnarray}
where we used the fact that $U|\phi\ra$ is also maximally entangled,
and the independence of $F_{\rm e}({\cal E})$ from the specific
maximally entangled state used in the definition.

%
%
Until now, our proof of Eq.~(\ref{eq:Horodecki}) has not deviated
substantially from~\cite{Horodecki99c}, and is included for
completeness.  The simplification is in the next step, namely, showing
that ${\cal E}_T$ is a depolarizing channel.  That is, there is a $p$
such that ${\cal E}_T(\rho) = pI/d + (1-p) \rho$ for all $\rho$.  The
proof of this fact in~\cite{Horodecki99c} made use of an isomorphism
between quantum operations and operators, while the following proof is
direct.  Note that for any unitary $V$,
\begin{eqnarray}
  V {\cal E}_T(\rho) V^{\dagger} = \int dU \,
  V U^{\dagger}{\cal E}( U\rho U^{\dagger} )
  UV^{\dagger}.
\end{eqnarray}
Making the change of variables $W \equiv UV^{\dagger}$ in the integral
we obtain
\begin{eqnarray} \label{eq:commutativity}
  V {\cal E}_T(\rho) V^{\dagger} = {\cal E}_T(V \rho V^{\dagger})
\end{eqnarray}
for all $\rho$ and $V$.  Let $P$ be a one-dimensional projector, and
$Q \equiv I-P$ be the projector onto the orthocomplementary space.
Letting $V$ be block diagonal with respect to the spaces onto which
$P$ and $Q$ project, we see that $V P V^{\dagger} = P$ and thus $V
{\cal E}_T(P) V^{\dagger} = {\cal E}_T(P)$.  It follows that ${\cal
  E}_T(P) = \alpha P + \beta Q$ for some $\alpha$ and $\beta$.  Using
$Q = I-P$, this expression may be rewritten as ${\cal E}_T(P) = p I/d
+ (1-p) P$, for some $p$, with $p$ possibly depending upon $P$.  Using
Eq.~(\ref{eq:commutativity}) again we see that this equation must hold
with the same value of $p$ for {\em any} one-dimensional projector
$P$.  By linearity of ${\cal E}_T$ it follows that ${\cal E}_T(\rho) =
pI/d +(1-p)\rho$ for all $\rho$, that is, ${\cal E}_T$ is a
depolarizing channel.

%
%
Finally, by direct calculation Eq.~(\ref{eq:Horodecki}) is easily
verified for depolarizing channels such as ${\cal E}_T$.  Since
$\overline{F}({\cal E}) = \overline{F}({\cal E}_T)$ and $F_{\rm
  e}({\cal E}) = F_{\rm e}({\cal E}_T)$ the result also holds for
general channels, which completes the proof.

%
%
Our next goal is to find a simple expression for $\overline{F}({\cal
  E})$ in terms of experimentally accessible quantities.  Let $\phi =
\sum_j |j\ra |j\ra / \sqrt{d}$ be a maximally entangled state of $RQ$.
Suppose we introduce a basis of unitary operators $U_j$ for a qudit,
with the $U_j$ orthogonal with respect to the Hilbert-Schmidt inner
product.  That is, $\tr(U_j^{\dagger} U_k) = \delta_{jk}d$, and thus
$U_j / \sqrt d$ forms an orthonormal operator basis.  An example of
such a set is operators of the form $X^kZ^l$ where the action of $X$
and $Z$ on computational basis states $|0\ra,\ldots,|d-1\ra$ is
defined by $X|j\ra \equiv |j\oplus 1\ra$, where $\oplus$ is addition
modulo $d$, and $Z|j\ra \equiv e^{2\pi i j/d}|j\ra$.  Other examples
of orthogonal unitary operator bases and general theory may be found
in~\cite{Knill96b,Knill96c,Klappenecker00a}.

%
%
Since $U_j/\sqrt d$ forms an orthonormal operator basis for a qudit,
$U_j^*/\sqrt d$ also forms an orthonormal operator basis, whence
$U_j^* \otimes U_k / d$ is an orthonormal operator basis for $RQ$.
It follows that
\begin{eqnarray}
  |\phi\ra \la \phi|
 = \sum_{jk} \frac{U_j^* \otimes U_k}{d} \frac{\tr\left( (U_j^* \otimes 
  U_k)^{\dagger} \phi\right)}{d}.
\end{eqnarray}
Note however that
\begin{eqnarray}
  \tr\left( (U_j^* \otimes U_k)^{\dagger} \phi\right) & = & \la \phi| U_j^T
 \otimes U_k^{\dagger} |\phi\ra 
 = \la \phi| I \otimes U_k^\dagger U_j |\phi\ra, \nonumber \\
 & & 
\end{eqnarray}
where we used the easily verified fact that $(A \otimes I) |\phi\ra =
(I \otimes A^T) |\phi \ra$.  Direct calculation shows that
\begin{eqnarray}
  \la \phi| I \otimes U_k^\dagger U_j |\phi\ra = 
\frac{\tr(U_k^{\dagger} U_j)}{d} = \delta_{jk}.
\end{eqnarray}
Substituting we obtain $\phi = \sum_{j} (U_j^* \otimes U_j)/d^2$.  It
follows that the entanglement fidelity is given by
\begin{eqnarray}
  F_{\rm e}({\cal E}) & = & \la \phi| {\cal E}(\phi) |\phi\ra 
  = \tr(\phi^\dagger {\cal E}(\phi) ) \\
  & = &  \frac{\sum_{jk} \tr\left((U_j^*)^\dagger U_k^* \otimes U_j^\dagger
      {\cal E}(U_k) \right)}{d^4} \\
  & = &  \frac{\sum_j \tr\left( U_j^{\dagger} {\cal E}(U_j) \right)}{d^3}.
\end{eqnarray}
(Compare also the related Eqs.~(6) and~(10) in~\cite{Fortunato02a},
which were obtained by different techniques, and which can also serve
as the basis for experimental determination of the entanglement
fidelity, and thus of the average fidelity, c.f. Eqs.~(17) and (18)
and the surrounding discussion in~\cite{Fortunato02b}.)  Using this
equation and Eq.~(\ref{eq:Horodecki}) we obtain the following formula
for the average gate fidelity
\begin{eqnarray} \label{eq:avg_fidelity}
\overline{F}({\cal E},U) = \overline{F}({\cal U^{\dagger}} \circ {\cal E})
 =\frac{\sum_j \tr \left( U U_j^{\dagger}U^{\dagger} {\cal E}(U_j)
    \right) + d^2}{d^2(d+1)}.
\end{eqnarray}
When $d = 2$ and choosing the $U_j$ to be the Pauli matrices $I, X, Y,
Z$ we obtain the result of~\cite{Bowdrey02a},
\begin{eqnarray}
  \overline{F}({\cal E},U) = \frac{1}{2} + \frac{1}{12} \sum_{j=1,2,3}
  \tr(U \sigma_j U^{\dagger} {\cal E}(\sigma_j)).
\end{eqnarray}

%
%
Eq.~(\ref{eq:avg_fidelity}) is theoretically interesting as a simple,
compact expression for the average gate fidelity, and may also be
interesting for experiment. Suppose one wished to experimentally
determine $\overline{F}({\cal E},U)$.  One way is to determine ${\cal
  E}$ directly via {\em quantum process
  tomography}~\cite{Chuang97a,Poyatos97a}, as demonstrated
in~\cite{Nielsen98b}, and then substitute into
Eq.~(\ref{eq:avg_fidelity}).  However, process tomography is complex
and its theoretical properties are not so easy to analyse.  A more
direct approach is to choose a set $\rho_k$ of quantum states which
form an operator basis, and which may be experimentally prepared with
high accuracy.  For example, such a set may be obtained from the
computational basis states $|0\ra,\ldots,|d-1\ra$ and superpositions
$(|j\ra \pm |k\ra)/\sqrt 2$, where $j \neq k$.  Many other sets of
states also suffice.  Standard linear algebraic methods may be used to
find co-efficients $\alpha_{jk}$ such that $U_j =\sum_k \alpha_{jk}
\rho_k$, whence Eq.~(\ref{eq:avg_fidelity}) implies
\begin{eqnarray}
\overline{F}({\cal E},U)  = \frac{\sum_{jk} \alpha_{jk}
  \tr \left( U U_j^{\dagger} U^{\dagger} {\cal E}(\rho_k)
    \right) + d^2}{d^2(d+1)}.
\end{eqnarray}
Using standard state tomography (see, e.g.~\cite{Leonhardt97a}) it is
possible to determine ${\cal E}(\rho_k)$, and thus to determine
$\overline{F}({\cal E},U)$.

%
%
In conclusion, we have obtained a simple formula for the average
fidelity of a noisy quantum channel or quantum gate.  This formula may
be useful for experimentally characterizing quantum gates and
channels.  It would be interesting to generalize these results further
to non-uniform starting distributions of states.

\acknowledgements Thanks to Jennifer Dodd, Gerard Milburn, Tobias
Osborne, and Lorenza Viola for their comments on the manuscript.


\begin{thebibliography}{13}
\expandafter\ifx\csname natexlab\endcsname\relax\def\natexlab#1{#1}\fi
\expandafter\ifx\csname bibnamefont\endcsname\relax
  \def\bibnamefont#1{#1}\fi
\expandafter\ifx\csname bibfnamefont\endcsname\relax
  \def\bibfnamefont#1{#1}\fi
\expandafter\ifx\csname citenamefont\endcsname\relax
  \def\citenamefont#1{#1}\fi
\expandafter\ifx\csname url\endcsname\relax
  \def\url#1{\texttt{#1}}\fi
\expandafter\ifx\csname urlprefix\endcsname\relax\def\urlprefix{URL }\fi
\providecommand{\bibinfo}[2]{#2}
\providecommand{\eprint}[2][]{\url{#2}}

\bibitem[{\citenamefont{Nielsen and Chuang}(2000)}]{Nielsen00a}
\bibinfo{author}{\bibfnamefont{M.~A.} \bibnamefont{Nielsen}} \bibnamefont{and}
  \bibinfo{author}{\bibfnamefont{I.~L.} \bibnamefont{Chuang}},
  \emph{\bibinfo{title}{Quantum computation and quantum information}}
  (\bibinfo{publisher}{Cambridge University Press},
  \bibinfo{address}{Cambridge}, \bibinfo{year}{2000}).

\bibitem[{\citenamefont{Schumacher}(1996)}]{Schumacher96a}
\bibinfo{author}{\bibfnamefont{B.~W.} \bibnamefont{Schumacher}},
  \bibinfo{journal}{Phys. Rev. A} \textbf{\bibinfo{volume}{54}},
  \bibinfo{pages}{2614} (\bibinfo{year}{1996}).

\bibitem[{\citenamefont{Bowdrey et~al.}(2002)\citenamefont{Bowdrey, Oi, Short,
  Banaszek, and Jones}}]{Bowdrey02a}
\bibinfo{author}{\bibfnamefont{M.~D.} \bibnamefont{Bowdrey}},
  \bibinfo{author}{\bibfnamefont{D.~K.~L.} \bibnamefont{Oi}},
  \bibinfo{author}{\bibfnamefont{A.~J.} \bibnamefont{Short}},
  \bibinfo{author}{\bibfnamefont{K.}~\bibnamefont{Banaszek}}, \bibnamefont{and}
  \bibinfo{author}{\bibfnamefont{J.~A.} \bibnamefont{Jones}},
  \bibinfo{journal}{Phys. Lett. A} \textbf{\bibinfo{volume}{294}},
  \bibinfo{pages}{258} (\bibinfo{year}{2002}),
  \bibinfo{note}{{arXive}:quant-ph/0201106}.

\bibitem[{\citenamefont{Fortunato
  et~al.}(2002{\natexlab{a}})\citenamefont{Fortunato, Pravia, Boulant,
  Teklemariam, Havel, and Cory}}]{Fortunato02a}
\bibinfo{author}{\bibfnamefont{E.~M.} \bibnamefont{Fortunato}},
  \bibinfo{author}{\bibfnamefont{M.~A.} \bibnamefont{Pravia}},
  \bibinfo{author}{\bibfnamefont{N.}~\bibnamefont{Boulant}},
  \bibinfo{author}{\bibfnamefont{G.}~\bibnamefont{Teklemariam}},
  \bibinfo{author}{\bibfnamefont{T.~F.} \bibnamefont{Havel}}, \bibnamefont{and}
  \bibinfo{author}{\bibfnamefont{D.~G.} \bibnamefont{Cory}},
  \bibinfo{journal}{J. Chem. Phys.} \textbf{\bibinfo{volume}{116}},
  \bibinfo{pages}{7599} (\bibinfo{year}{2002}{\natexlab{a}}),
  \bibinfo{note}{{arXiv}:quant-ph/0202065}.

\bibitem[{\citenamefont{Fortunato
  et~al.}(2002{\natexlab{b}})\citenamefont{Fortunato, Viola, Hodges,
  Teklemariam, and Cory}}]{Fortunato02b}
\bibinfo{author}{\bibfnamefont{E.~M.} \bibnamefont{Fortunato}},
  \bibinfo{author}{\bibfnamefont{L.}~\bibnamefont{Viola}},
  \bibinfo{author}{\bibfnamefont{J.}~\bibnamefont{Hodges}},
  \bibinfo{author}{\bibfnamefont{G.}~\bibnamefont{Teklemariam}},
  \bibnamefont{and} \bibinfo{author}{\bibfnamefont{D.~G.} \bibnamefont{Cory}},
  \bibinfo{journal}{New J. Phys.} \textbf{\bibinfo{volume}{4}},
  \bibinfo{pages}{5.1} (\bibinfo{year}{2002}{\natexlab{b}}),
  \bibinfo{note}{{arXiv}:quant-ph/0111166}.

\bibitem[{\citenamefont{Horodecki et~al.}(1999)\citenamefont{Horodecki,
  Horodecki, and Horodecki}}]{Horodecki99c}
\bibinfo{author}{\bibfnamefont{M.}~\bibnamefont{Horodecki}},
  \bibinfo{author}{\bibfnamefont{P.}~\bibnamefont{Horodecki}},
  \bibnamefont{and}
  \bibinfo{author}{\bibfnamefont{R.}~\bibnamefont{Horodecki}},
  \bibinfo{journal}{Phys.~Rev.~A} \textbf{\bibinfo{volume}{60}},
  \bibinfo{pages}{1888} (\bibinfo{year}{1999}).

\bibitem[{\citenamefont{Knill}(1996{\natexlab{a}})}]{Knill96b}
\bibinfo{author}{\bibfnamefont{E.}~\bibnamefont{Knill}},
  \bibinfo{journal}{{arXiv}:quant-ph/9608048}
  (\bibinfo{year}{1996}{\natexlab{a}}).

\bibitem[{\citenamefont{Knill}(1996{\natexlab{b}})}]{Knill96c}
\bibinfo{author}{\bibfnamefont{E.}~\bibnamefont{Knill}},
  \bibinfo{journal}{{arXiv}:quant-ph/9608049}
  (\bibinfo{year}{1996}{\natexlab{b}}).

\bibitem[{\citenamefont{Klappenecker and Roetteler}(2000)}]{Klappenecker00a}
\bibinfo{author}{\bibfnamefont{A.}~\bibnamefont{Klappenecker}}
  \bibnamefont{and}
  \bibinfo{author}{\bibfnamefont{M.}~\bibnamefont{Roetteler}},
  \bibinfo{journal}{{arXiv}:quant-ph/0010082}  (\bibinfo{year}{2000}).

\bibitem[{\citenamefont{Chuang and Nielsen}(1997)}]{Chuang97a}
\bibinfo{author}{\bibfnamefont{I.~L.} \bibnamefont{Chuang}} \bibnamefont{and}
  \bibinfo{author}{\bibfnamefont{M.~A.} \bibnamefont{Nielsen}},
  \bibinfo{journal}{J. Mod. Opt.} \textbf{\bibinfo{volume}{44}},
  \bibinfo{pages}{2455} (\bibinfo{year}{1997}),
  \bibinfo{note}{{arXiv}:quant-ph/9610001}.

\bibitem[{\citenamefont{Poyatos et~al.}(1997)\citenamefont{Poyatos, Cirac, and
  Zoller}}]{Poyatos97a}
\bibinfo{author}{\bibfnamefont{J.~F.} \bibnamefont{Poyatos}},
  \bibinfo{author}{\bibfnamefont{J.~I.} \bibnamefont{Cirac}}, \bibnamefont{and}
  \bibinfo{author}{\bibfnamefont{P.}~\bibnamefont{Zoller}},
  \bibinfo{journal}{Phys. Rev. Lett.} \textbf{\bibinfo{volume}{78}},
  \bibinfo{pages}{390} (\bibinfo{year}{1997}).

\bibitem[{\citenamefont{Nielsen et~al.}(1998)\citenamefont{Nielsen, Knill, and
  Laflamme}}]{Nielsen98b}
\bibinfo{author}{\bibfnamefont{M.~A.} \bibnamefont{Nielsen}},
  \bibinfo{author}{\bibfnamefont{E.}~\bibnamefont{Knill}}, \bibnamefont{and}
  \bibinfo{author}{\bibfnamefont{R.}~\bibnamefont{Laflamme}},
  \bibinfo{journal}{Nature} \textbf{\bibinfo{volume}{396}}, \bibinfo{pages}{52}
  (\bibinfo{year}{1998}).

\bibitem[{\citenamefont{Leonhardt}(1997)}]{Leonhardt97a}
\bibinfo{author}{\bibfnamefont{U.}~\bibnamefont{Leonhardt}},
  \emph{\bibinfo{title}{Measuring the quantum state of light}}
  (\bibinfo{publisher}{Cambridge University Press}, \bibinfo{address}{New
  York}, \bibinfo{year}{1997}).

\end{thebibliography}

\end{document}